\documentclass[prd,superscriptaddress,showpacs,preprintnumbers,amsmath,amssymb]{revtex4}
\usepackage{graphicx}% Include figure files
\usepackage{dcolumn}% Align table columns on decimal point
\usepackage{bm}% bold math
\usepackage{epsfig}
\usepackage{amsfonts}

\begin{document}

\title{Massive uncharged and charged particles' tunneling from the
Horowitz-Strominger Dilaton black hole }

\author{Yapeng Hu}
  \email{huzhengzhong2050@163.com}
  \affiliation{Department of Physics,
Beijing Normal University, Beijing 100875, China}

\author{Jingyi Zhang}
   \email{physicz@263.net}
   \affiliation {Center for Astrophysics,
Guangzhou University, Guangzhou 510006, China}

\author{Zheng Zhao}
   \email{zhaoz43@hotmail.com}
   \affiliation {Department of Physics,
Beijing Normal University, Beijing 100875, China}

\begin{abstract}
Originally, Parikh and Wilczek's work is only suitable for the massless
particles' tunneling. But their work has been further extended to the cases
of massive uncharged and charged particles' tunneling recently. In this
paper, as a particular black hole solution, we apply this extended method to
reconsider the tunneling effect of the H.S Dilaton black hole. We
investigate the behavior of both massive uncharged and charged particles,
and respectively calculate the emission rate at the event horizon. Our
result shows that their emission rates are also consistent with the unitary
theory. Moreover, comparing with the case of massless particles' tunneling,
we find that this conclusion is independent of the kind of particles. And it
is probably caused by the underlying relationship between this method and
the laws of black hole thermodynamics.

\end{abstract}
 \maketitle

\section{Introduction}

Since Parikh and Wilczek gave a new viewpoint on the Hawking radiation in
2000\cite{1,2,3,4}, their original works have been extended to many black
holes. However, all the particles that these works researched are massless,
and the space-times are static\cite{5,6,7,8,9,10,11,12,13}. Recently, Parikh
and Wilczek's original method has been further extended to the following
cases-the stationary space-times, massive particles' tunneling and even the
charged particles' tunneling\cite{14,15,16,17,18}. Furthermore, these works'
results are also consistent with the underlying unitary theory. In this
paper, as a particular black hole solution, a spherically symmetric charged
dilaton black hole which is from the string theory, we reconsider the
tunneling effect of the H.S Dilaton black hole\cite{13,21}. As we know,
usually, the 3+1 dimensional charged dilaton black hole is obtained from the
low energy effective field theory which describes strings, and it has many
properties different from those black holes in ordinary Einstein gravity\cite%
{19,20,22,23} (which can be simply seen from that the calculation is more
complex in Ref\cite{13,21}). Thus, using this extended method, we can
reinvestigate the behavior of both the massive uncharged and charged
particles tunneling across the event horizon of the H.S Dilaton black hole,
and then calculate their corresponding emission rates. Different from the
massless particles, the massive particles don't follow the radial lightlike
geodesics when they tunnel across the event horizon. And even, the action
for the classical forbidden trajectory should be modified for the massive
charged particles, because of the existence of the electromagnetic field
outside the H.S Dilaton black hole. However, if we consider the massive
tunneling particle as a massive shell (de Broglie s-wave), we can
investigate the behavior of both the massive uncharged and charged particles
when they tunnel across the event horizon\cite{14,15,18}.

The rest of the paper is organized as follows. In section 2, we introduce
the Painleve-H.S Dilaton coordinate system, and investigate the behavior of
massive particles tunneling across the event horizon\cite{13,21,24}. In
section 3, first, we calculate the emission rate of massive particles, and
then after modifying the action for the classical forbidden trajectory, we
also calculate the emission rate of massive charged particles, and obtain
their correct spectrums. In section 4, we give a conclusion and discussion
on our result.

\section{Painleve-H.S coordinates and the behavior of the massive tunneling
particles}

The line element of the Horowitz-Strominger black hole is\cite%
{13,19,20,21,22,23}

\begin{equation}
ds^{2}=(1-\frac{r_{H}}{r})(1-\frac{r_{-}}{r})^{\frac{1}{1+P}}dt_{h}^{2}-%
\frac{1}{(1-\frac{r_{H}}{r})(1-\frac{r_{-}}{r})^{\frac{1}{1+P}}}%
dr^{2}-R^{2}(r)(d\theta ^{2}+\sin ^{2}\theta d\varphi ^{2}).  \label{1}
\end{equation}%
where $r_{H}$, $r_{-}$ are respectively the event horizon and the inner
horizon, $P$ is a constant, and $R(r)=r(1-\frac{r_{-}}{r})^{\frac{P}{2(P+1)}%
} $. In addition, the mass and the charge of the black hole are respectively

\begin{eqnarray}
M &=&\frac{r_{H}}{2}+\frac{1}{1+P}\frac{r_{-}}{2},  \notag \\
Q^{2} &=&\frac{r_{H}r_{-}(2+P)}{2(1+P)}.  \label{2}
\end{eqnarray}

In fact, if $r_{H}<r_{-}$, there will not even be a black hole. Thus, we set
$r_{H}>r_{-}$ . And the Painleve-H.S Dilaton black hole metric is

\begin{eqnarray}
ds^{2} &=&(1-g)\Delta dt^{2}-2\sqrt{1-(1-g)\Delta }dtdr-dr^{2}-R^{2}(r)(d%
\theta ^{2}+\sin ^{2}\theta d\varphi ^{2})  \notag \\
&=&g_{00}dt^{2}+2g_{01}dtdr-dr^{2}-R^{2}(r)(d\theta ^{2}+\sin ^{2}\theta
d\varphi ^{2}).  \label{3}
\end{eqnarray}%
which can be obtained from the H.S Dilaton line element (1) by the
coordinate transformation

\begin{equation}
dt=dt_{h}+\frac{1}{\Delta }\frac{\sqrt{1-(1-g)\Delta }}{1-g}dr.  \label{4}
\end{equation}
where \

\begin{equation}
g=\frac{r_{H}}{r},\Delta =(1-\frac{r_{-}}{r})^{\frac{1}{1+P}}.  \label{5}
\end{equation}

It is easy to find that the metric (3) displays the stationary, nonstatic,
and nonsingular nature of the space time at the event horizon $r_{H}$.
Moreover, it satisfies the Landau's condition of the coordinate clock
synchronization\cite{25,26}. All these features are available for us to
investigate the behavior of the tunneling massive particles, and which will
be seen as follows.

In Parikh and Wilczek's original works, a significant key-point is to
calculate the imaginary part of the action, and the first is to investigate
the behavior of the tunneling particles in the corresponding Painleve
coordinates\cite{1,2,3,4}. As before, the behavior of the tunneling massless
particles can be given by the radial null geodesics. However, the massive
particles don't follow the null geodesic when they tunnel across the event
horizon, and even the massive charged particles will be the subject to
Lorentz forces. In order to investigate the behavior of the tunneling
massive particles, we can consider both the outgoing massive uncharged and
outgoing massive charged particle as a massive shell (non-relativistic de
Broglie s-wave). The difference is that the massive shell is uncharged for
the massive uncharged particle, while the massive shell is charged for the
massive charged particle\cite{14,15,18}. On the other hand, according to the
non-relativistic quantum mechanics describing the massive particles'
tunneling, the de Broglie's hypothesis and the WKB formula, we can find that
the behavior of the tunneling massive particles can be approximately
determined by the phase velocity of the de Broglie s-wave whose phase
velocity $v_{p}$ and group velocity $v_{g}$ have the following relationship%
\cite{14,15,18}

\begin{equation}
v_{p}=\frac{1}{2}v_{g},v_{p}=\frac{dr}{dt}=\overset{\cdot }{r},v_{g}=\frac{%
dr_{c}}{dt}.  \label{6}
\end{equation}%
where $r_{c}$ is the radial position of the particle.

Since tunneling across the barrier is an instantaneous process, there are
two events that take place simultaneously in different places during the
process. One event is massive particle tunneling into the barrier, and the
other is massive particle tunneling out the barrier. Because the metric (3)
satisfies the Landau's condition of the coordinate clock synchronization,
the difference of coordinate times of these two simultaneous events is

\begin{equation}
dt=-\frac{g_{0i}}{g_{00}}dx^{i}=-\frac{g_{01}}{g_{00}}dr_{c}\ \text{\ }%
(d\theta =d\varphi =0),  \label{7}
\end{equation}%
Thus, according to the relationship (6), the group velocity is

\begin{equation}
v_{g}=\frac{dr_{c}}{dt}=-\frac{g_{00}}{g_{01}},  \label{8}
\end{equation}%
and the phase velocity is

\begin{equation}
v_{p}=\overset{\cdot }{r}=\frac{1}{2}v_{g}=-\frac{1}{2}\frac{g_{00}}{g_{01}}=%
\frac{1}{2}\frac{(1-g)\Delta }{\sqrt{1-(1-g)\Delta }}.  \label{9}
\end{equation}%
Which is the behavior of both the outgoing massive uncharged and charged
particles. The difference between them would be indicated if we take the
self-interaction effect into consideration. Because when the particles
tunnel across the event horizon of the H.S Dilaton black hole, the mass and
charge in (3) and (9) should be respectively changed $M$ with $M-\omega $\ ,
and $Q$ with\ $Q-q$, where $\omega $ and $q$ are respectively the mass and
the charge of the tunneling particle.

\section{Tunneling rate}

\subsection{Massive uncharged particles' tunneling rate}

According to Parikh and Wilczek's original works, the tunneling rate $\Gamma
$ could take the following form\cite{2,27}

\begin{equation}
\Gamma \sim \exp (-2\text{Im}S).  \label{10}
\end{equation}%
and after using the Hamilton's equation $\frac{dH}{dp_{r}}=\overset{\cdot }{r%
}$, the imaginary part of the action is

\begin{equation}
\text{Im}S=\text{Im}\int_{r_{i}}^{\ r_{f}}p_{r}dr=\text{Im}%
\int\nolimits_{r_{i}}^{\ r_{f}}\int_{0}^{p_{r}}dp_{r}^{^{\prime }}dr=\text{Im%
}\int\nolimits_{r_{i}}^{\ r_{f}}\int_{M_{i}}^{M_{f}}\frac{dM}{\overset{\cdot
}{r}}dr.  \label{11}
\end{equation}%
Where $M_{i}$ and $M_{f}$ are respectively the initial mass and the final
mass of the black hole.

Thus, keeping $Q$ a constant, substituting (9) into (11), and then doing the
$r$ integral first, we obtain

\begin{equation}
\text{Im}S=\text{Im}\int_{M_{i}}^{M_{f}}\int\nolimits_{r_{i}}^{\ r_{f}}\frac{%
2\sqrt{r^{2}-(r-r_{H})r\Delta }drdM}{(r-r_{H})(1-\frac{r_{-}}{r})^{\frac{1}{%
1+P}}}=-\pi \int_{M_{i}}^{M_{f}}\frac{2r_{H}{}}{(1-\frac{r_{-}}{r_{H}})^{%
\frac{1}{1+P}}}dM.  \label{12}
\end{equation}%
\ Not as before, we don't integral (12) directly. Instead, we first
calculate the differential of the entropy $S$ to the mass $M$. And the
entropy of the H.S Dilaton black hole is\cite{28}

\begin{eqnarray}
S &=&\frac{1}{4}A=\pi R^{2}(r_{H})=\pi r_{H}^{2}(1-\frac{r_{-}}{r_{H}})^{%
\frac{P}{1+P}}  \notag \\
&=&\pi \lbrack r_{H}^{2}-Q^{2}\frac{2(1+P)}{2+P}]^{\frac{P}{1+P}}r_{H}^{%
\frac{2}{1+P}}.  \label{13}
\end{eqnarray}%
Therefore, the differential of the entropy $S$ to the mass $M$ is

\begin{equation}
dS=\frac{dS}{dr_{H}}\frac{dr_{H}}{dM}dM=\frac{4\pi r_{H}}{(1-\frac{r_{-}}{%
r_{H}})^{\frac{1}{1+P}}}dM.  \label{14}
\end{equation}%
From (14) and (12), we can easily obtain

\begin{equation}
\Gamma \sim e^{-2\text{Im}S}=e^{\Delta S}.  \label{15}
\end{equation}%
which is consistent with the underlying unitary theory and support the
conservation of information.

\subsection{Massive charged particles' tunneling rate}

When the massive charged particles tunnel across the event horizon of the
H.S Dilaton black hole, not only the mass but also the charge of black hole
will change for the conservation of energy and charge. In addition,
different from the massive uncharged particles, for the existence of the
electromagnetic filed outside the H.S Dilaton black hole, we should also
take its effect into account when the massive charged particles tunnel
across the event horizon. Therefore, we must consider the black hole and the
electromagnetic filed outside black hole as a whole matter-gravity system%
\cite{14,15,18}. For the H.S Dilaton black hole, the 4-dimensional
electromagnetic potential is

\begin{equation}
A_{\mu }=(A_{t},0,0,0).  \label{16}
\end{equation}%
where $A_{t}=-Q/r$. And the Lagrangian function of the electromagnetic filed
corresponding to the generalized coordinates described by $A_{\mu }$ is $%
L_{e}=-1/4F_{\mu \upsilon }F^{\mu \upsilon }$. Thus, when a massive charged
particle tunnels out, the whole system will transit from one state to
another. However, we find that the generalized coordinate $A_{t}$ is an
ignorable coordinate. In order to obtain the correct kinematical equation
from the differential of the action, we should eliminate the freedom
corresponding to $A_{t}$. Therefore, the action of the charged massive
particle should be written as

\begin{eqnarray}
S &=&\int_{t_{i}}^{t_{f}}(P_{r}\overset{\cdot }{r}-P_{A_{t}}\overset{\cdot }{%
A_{t}})dt  \notag \\
&=&\int_{r_{i}}^{r_{f}}[\int_{(0,0)}^{(P_{r},P_{A_{t}})}(\overset{\cdot }{r}%
dP_{r}^{\prime }-\overset{\cdot }{A_{t}}dP_{A_{t}}^{\prime })]\frac{dr}{%
\overset{\cdot }{r}}.  \label{17}
\end{eqnarray}

According to the Hamilton's equations, we have

\begin{eqnarray}
\overset{\cdot }{r} &=&\frac{dH}{dP_{r}}\mid _{(r;A_{t},P_{A_{t}})}=\frac{dM%
}{dP_{r}},  \notag \\
\overset{\cdot }{A_{t}} &=&\frac{dH}{dP_{A_{t}}}\mid _{(A_{t};r,P_{r})}=%
\frac{Q}{r}dQ.  \label{18}
\end{eqnarray}%
Substituting (18) into (17), and switching the order of integration yield
the imaginary part of the action

\begin{equation}
\text{Im}S=\text{Im}\int\nolimits_{r_{i}}^{\ r_{f}}\int_{(M,Q)}^{(M-\omega
,Q-q)}[dM-\frac{Q}{r}dQ]\frac{dr}{\overset{\cdot }{r}}.  \label{19}
\end{equation}%
We would like to emphasize that, since we consider the massive charged
particles' tunneling, the $Q$ in the expression of $\overset{\cdot }{r}$ in
(9) should also be changed. In addition, we also can find that the
conservation of energy and electric charge will be enforced in a natural
way. Thus, substituting the (9) into (19), and doing the $r$ integral first,
we obtain

\begin{eqnarray}
\text{Im}S &=&\text{Im}\int\nolimits_{r_{i}}^{\
r_{f}}\int_{(M,Q)}^{(M-\omega ,Q-q)}[dM-\frac{Q}{r}dQ]\frac{2\sqrt{%
r^{2}-(r-r_{H})r\Delta }dr}{(r-r_{H})(1-\frac{r_{-}}{r})^{\frac{1}{1+P}}}
\notag \\
&=&-\pi \int_{(M,Q)}^{(M-\omega ,Q-q)}[\frac{2r_{H}{}}{(1-\frac{r_{-}}{r_{H}}%
)^{\frac{1}{1+P}}}dM-\frac{2Q}{(1-\frac{r_{-}}{r_{H}})^{\frac{1}{1+P}}}dQ].
\label{20}
\end{eqnarray}%
Similarly, we don't directly calculate the integral (20). We first calculate
the differential of the entropy $S$ to both the mass $M$ and the electric
charge $Q$. According to the expression of the entropy in (13), the complete
differential of the entropy is

\begin{eqnarray}
dS &=&\frac{\partial S}{\partial r_{H}}\frac{\partial r_{H}}{\partial M}dM+(%
\frac{\partial S}{\partial r_{H}}\frac{\partial r_{H}}{\partial Q}+\frac{%
\partial S}{\partial Q})dQ  \notag \\
&=&\frac{4\pi r_{H}{}}{(1-\frac{r_{-}}{r_{H}})^{\frac{1}{1+P}}}dM-\frac{4\pi
Q}{(1-\frac{r_{-}}{r_{H}})^{\frac{1}{1+P}}}dQ.  \label{21}
\end{eqnarray}%
Comparing (21) with (20), we can also easily obtain the equation (15) and
the same conclusion, and find that the result takes the same functional form
as that of the massive uncharged particles.

\section{Conclusion and discussion}

Recently, the charged dilaton black holes have been much researched. As is
mentioned in section one, the charged dilaton black holes have many
properties\ different from those black holes in ordinary Einstein gravity.
Thus, in this paper, using the extended method\cite{14,15,18}, we first
investigate the behavior of both the massive uncharged and charged particles
which tunnel across the event horizon of the H.S charged dilaton black hole,
and then reconsider the tunneling effect of it. But our result shows that
their emission rates are also consistent with the unitary theory, the same
as that of the case of massless particles' tunneling, which manifests that
the conclusion is independent of the kind of particles. On the other hand,
as is seen from the Refs\cite{13,21}, for the massless particles' tunneling
of the H.S Dilaton black hole, the authors do directly calculate the
imaginary part of the action after doing the $r$ integral, and this direct
calculation is usually complex. However, in our paper, we simplify the
calculation. That, after doing the $r$ integral, we don't directly calculate
it. Instead, we calculate the differential of the entropy $S$ to the
variance, compare it with the imaginary part of the action after doing the $r
$ integral, and then we can easily obtain the same equation (15) and
conclusion. But why is the conclusion independent of the kind of particles?
does this two sides have relationship? In the following part, we will give a
discussion on these two questions. Our discussion shows that, in viewing of
the laws of black hole thermodynamics, these two questions may be solved
well, which is explained in detail in the following.

As we know, according to the first law of black hole thermodynamics, the
general differential Bekenstein-Smarr equation of the black hole is\cite{29}

\begin{equation}
dM=\frac{\kappa }{8\pi }dA+VdQ+\Omega dJ  \label{22}
\end{equation}%
Therefore, if we consider that the tunnelling process is a reversible
process, according to the second law of black hole thermodynamics, (22) can
be rewritten as

\begin{equation}
dM=TdS+VdQ+\Omega dJ  \label{23}
\end{equation}%
which is equivalent to

\begin{equation}
dS=\frac{dM}{T}-\frac{VdQ}{T}-\frac{\Omega dJ}{T}  \label{24}
\end{equation}

In our cases, we investigate the behavior of the particles tunneling across
the event horizon of the H.S Dilaton black hole. The temperature $T$ and the
electro-potential $V$ are\cite{20}

\begin{equation}
T=\frac{1}{4\pi }(1-\frac{r_{-}}{r_{+}})^{\frac{1}{1+P}},V=\frac{Q}{r}.
\label{25}
\end{equation}

Thus, when we consider the massive uncharged or charged particles tunneling
across the event horizon without the angle momentum, the equations
corresponding to (24) are just respectively the (14) and the (21). Moreover,
in Refs\cite{13,21}, we can also obtain the corresponding equation from
(24). It implies that not only our simplified calculation has hidden the
laws of black hole thermodynamics, but also all those methods (either the
original Parikh and Wilczek's method or its extended method) have the
underlying relationship with the laws of black hole thermodynamics. As we
know, the laws of black hole thermodynamics are independent of the kind of
tunneling particles, thus, the independence of the conclusion may come from
it. (Note that, we had already found the conclusion's independence of the
kind of particles in the Ref\cite{30}. And concerning this independence, we
gave a brief discussion. However, this discussion is not complete. Thus,
this paper can also be considered as the advanced discussion about the
independence).

\section{Acknowledgements}

This work is supported by the National Natural Science Foundation of
China under Grant No.10475013, No.10373003 and the National Basic
Research Program of China Grant No.2003CB716300.

\end{document}